\documentclass[12pt,a4paper]{article}
\usepackage{cite}
\newcommand{\sfrac}[2]{{\textstyle\frac{#1}{#2}}}

\def\joinrel{\mathrel{\mkern-4mu}}
\newcommand{\xarr}{\relbar\joinrel}
\newcommand{\arrowcond}[2]{\mathrel{\mathop{%
\xarr\xarr\xarr\rightarrow}\limits^{\!#1}_{\!#2}}}
\begin{document}

\title{
The Vortex Solution in the (2+1)-Dimensional 
Yang-Mills-Chern-Simons Theory at High Temperature
}

\author{
V.V.Skalozub\thanks{e-mail address: skalozub@ff.dsu.dp.ua}\ \  
and  
A.Yu.Zaslavsky\thanks{e-mail address: alex@ff.dsu.dp.ua} 
\\
{\em Dniepropetrovsk State University} \\ 
{\em Dniepropetrovsk, 320625 Ukraine } 
}

\date{5th May 1999}

\maketitle

\begin{abstract}
The vortex-like solution to the no-linear field equations in a 
two-dimensional $SU(2)$ gauge theory with the Chern-Simons mass term is 
found at high temperature. It is derived from the effective Lagrangian 
including the leading order finite temperature corrections. The discovered 
field configuration possesses the finite energy and the quantized magnetic 
flux. At the centre of the vortex the point charge is located which is 
surrounded by the distributed charge of the opposite sign and the vortex 
is neutral as a whole. At high temperature the energy of the vortex is 
negative and it corresponds to the ground state. The derived solution is 
considered to be a result of heating the lattice vacuum structure formed 
at zero temperature.
\end{abstract}

\section{Introduction}

Over the last decade gauge theories in two spatial dimensions have 
been the object of considerable interest. The origins of this interest 
are manifold. 
First, these theories have a number of interesting 
features which make them quite different from theories in the conventional 
space-time. There is, for example, a possibility of introducing 
a mass term for the gauge field---the Chern-Simons (CS) topological 
mass---in a gauge invariant way~\cite{DJT:82}. In fact, if not added by 
hand, this mass is necessarily generated by the fermion loop 
corrections~\cite{Redl:84,NieS:83}.
Second, at high temperature field theories in a (3+1)-dimensional 
space-time effectively behave as 
theories in a 3-dimensional Euclidean space~\cite{GPY:81}.
Third, low-dimensional theories are successfully applied to the condensed 
matter physics phenomena---quantum Hall effect, high-T 
superconductivity---which proved to have an essentially 
planar nature~\cite{KrR:87e,RDSS:90}. 

It is well known that the non-Abelian gauge field vacuum becomes 
unstable in an external magnetic field due to a tachyonic mode present
in the gluon spectrum~\cite{Sk:78e,NilOl:78}. This holds for both
three- and two-dimensional models. Searches for the true vacuum state 
of the non-abelian gauge fields 
have led to the discovery of gauge field condensation in an external 
magnetic field~\cite{Sk:86e,AmO:90,MDT:92}. 
The mechanism of the structure generation included exciting of 
tachyonic modes in intense magnetic fields. Also, a scalar field 
should be present in order to stabilize the condensate. 
An essential feature of this vacuum is its lattice structure which 
results in a breakdown of vacuum homogeneity. This picture is similar 
to that of superconductivity~\cite{Abr:57e}, although the origin of 
tachyonic modes is different.  

In Ref.~\cite{SVZ:94pr} it has been shown that in 2D gauge theories 
there is a mechanism of the vacuum structure formation quite distinct 
from that in 3+1 space-time. 
It occurs due to the CS mass $m$ presence, in the magnetic fields below 
the tachyonic threshold in the gluon spectrum, and does not 
involve the condensation of unstable modes. 
The vacuum state of the $SU(2)$ theory in external magnetic field has 
been found to be a periodic electromagnetic structure; 
both magnetic and electric fields are present, and form a 
triangular lattice with a period $\sim 1/m$. 
Each cell contains magnetic flux quantized in a manner similar to
superconductivity.
The configuration is a perturbative solution to the non-linear field 
equations. 
Generation of a pure magnetic structure is suppressed by the CS term, 
which mixes electric and magnetic components of the colour 
fields. 
As a result, the $A_0$ component of the gauge potential plays an important
role in the vacuum structure formation.
As the threshold of the tachyonic mode appearance set by the topological 
mass is not reached, the derived state is stable. 
The lattice has been found to lower the energy of the applied homogeneous 
magnetic field. 
Moreover, when $H=0$, the energy density is negative and the condensate 
can be generated spontaneously without an external magnetic field.  
It is the CS mass that gives the origin to this new type of vacuum 
structure, which has no analogues among lattices studied in the other 
models. 

The analysis carried out in Ref.~\cite{SVZ:94pr} has a general character 
and allows extension to the case when the CS mass is generated at the 
one-loop level. 
Therefore, the noted mechanism could produce the gluon field 
vacuum structure in various external environments. 
This possibility has been examined in the finite temperature 
model~\cite{SZ:98e}. 
The technique applied was essentially the one of zero temperature case. 
As it appeared, the periodic vacuum structure persists at finite 
temperature, and is generated due to the same mechanism as at $T=0$. 
The crucial factor for the lattice formation is again the topological mass 
presence.
However, it has been found that the amplitude of the gauge field 
condensate increases as the temperature rises. 
As a consequence, at high temperature $T\gg m$ the results obtained  
within a perturbative method should be treated with 
caution. 
It has been argued that the analysis of the non-linear 
equations is needed to describe the vacuum evolution at high 
temperature.

The main goal of the present paper is to investigate how the vacuum 
structure is affected by high temperature when the perturbation
theory is not adequate. To do this we solve the 
non-linear equations of motion for the topologically massive gluon 
fields derived from the effective Lagrangian (EL). 
In contrast to the magnetic lattice case, when the CS mass is present
an electric field is generated in a vacuum and, so, the $A_0$ 
component is not decoupled. Therefore, we should not fail to take into 
account the Debye mass when constructing the ${\cal L}_{eff}$.
We consider the one-loop gluon effective Lagrangian
of the two-dimensional gauge theory at high temperature.
The thermal contribution is introduced 
via the one-loop gluon polarization tensor in the static limit. 
We take into consideration the leading in $T$ term of the Debye mass,
$\mu$, which is $\sim g^2T$ in 2+1 dimensions. 
The one-loop induced CS coefficient in the EL is usually modified at 
finite temperature by the factor of $\tanh m_f/T$. We will discuss 
the implications of such temperature dependence below. 
With thermal effects thus included, the following results have been 
obtained. 

An exact solution to the non-linear gluon field equations has been 
found. The solution is static and axially symmetric; the fields
form a vortex in both coordinate and gauge spaces. 
At spatial boundary the gluon potentials constituting the solution 
become a pure gauge. 
Both electric and magnetic fields are present in the derived
configuration. 
Due to the CS and the temperature masses presence the fields are 
short-range, and vanish exponentially at $r\to\infty$ and $T\to\infty$. 
The extent of the area in which the fields are localized is determined 
by the inverted effective mass $M=\sqrt{m^2+\mu^2}$. 
So, at spatial infinity the configuration approaches
the perturbative vacuum. The important feature of the found solution is
finiteness of the total energy of the vortex. 
The sign of energy is not constant, it is determined by the sign of the
combination $m^2-\mu^2$. 
The electric field has a Coulomb-like behaviour at the origin signalling 
the presence of a point charge. It cancels exactly the density of the 
continuously distributed charge resulting in the neutral field 
configuration.
Thus, the derived solution to the field equation describes the 
composite of a vortex, which possesses finite energy 
and magnetic flux, and a point charge.

The picture summarized above corresponds quite well to the lattice 
solution found in Refs.~\cite{SVZ:94pr,SZ:98e}. 
In fact, the vortex can be considered as a result of heating the 
vacuum structure. Depending on the sign of the free energy, 
the vortex solution can describe either particle-like excitation
over the periodic vacuum structure, or the vacuum state 
represented by the system of isolated vortices.
We note once again that the discovered configuration exists because the
equation of motion are derived from the effective Lagrangian, which
includes the temperature mass contribution for $A_0$ component of the 
gluon field.

\section{Vortex solution to the field equations}

Two-dimensional $SU(2)$ gauge theory in the Minkowski space-time is 
governed by the following generating functional:
\begin{equation}\label{action} 
Z=\int d\psi\,d{\bar\psi}\,dA 
\exp\left[i\int d^3x\,\Bigl[-\sfrac{1}{4} G^{\mu\nu}_aG_{\mu\nu}^a + 
{\bar\psi}\gamma^\mu(i\partial_\mu + 
\sfrac{1}{2}g\sigma_aA^a_\mu+m_f)\psi\Bigr]\right], 
\end{equation}
where 
$G^{a}_{\mu\nu}=\partial_{\mu}A^{a}_{\nu}-\partial_{\nu}A^{a}_{\mu}+
g\varepsilon_{abc}A^{b}_{\mu}A^{c}_{\nu}$, 
$A^{a}_{\mu}$ are Yang-Mills (YM) potentials, 
$\psi$, $\bar\psi$ are the fermion fields, 
$\sigma_a$ are Pauli matrices,
$g$ is the gauge coupling constant. 

We want to build the one loop effective Lagrangian of gauge fields
and, to do this, 
integrate over the fermion degrees of freedom. As a result,  
two additional terms appear in ${\cal L}_{eff}$. The first is 
contributed by the parity-even part of gluon polarization tensor 
$\Pi_{\mu\nu}^{ab}$; 
the second is parity-violating one intrinsic to odd-dimensional
space-time, the well known CS term~\cite{Redl:84,NieS:83}:  
\begin{equation}\label{Lcs} 
  {\cal L}_{CS}={\textstyle\frac{1}{4}}m\varepsilon^{\alpha\beta\gamma} 
    \Bigl(G^{a}_{\alpha\beta}
   A^{a}_{\gamma}-{\textstyle\frac{1}{3}}g\varepsilon_{abc}
    A^{a}_{\alpha}
   A^{b}_{\beta}A^{c}_{\gamma}\Bigr),
\end{equation}
where $m$ is real. This term renders the gauge field excitations
massive, with the mass~$m$. For the theory to remain invariant under 
large gauge transformations, the CS coefficient must be quantized.
To be precise, the value of $m$ should satisfy the condition
\begin{equation}\label{quam}
\frac{m}{g^2} = \frac{n}{4\pi}.
\end{equation}
where $n$ is an integer. We note that the coupling $g$ has dimension 
of $m^{1/2}$ in 2+1 space-time.

We are interested in obtaining the gluon effective Lagrangian at finite 
temperature. The polarization tensor 
$\Pi_{\mu\nu}^{ab} = \Pi_{\mu\nu}\delta^{ab}$ 
contribution has been obtained
by the authors earlier~\cite{SZ:98e} in high-temperature approximation.
The following expression for the $\Pi_{00}$ component has been calculated
\begin{equation}\label{pi00} 
\Pi_{00}(T,k_0=0,\vec{k}\to 0) \equiv \mu^2(T) = 
g^2T\biggl(\frac{\ln 2}{2\pi} - \frac{1}{16\pi}\frac{m_f^2}{T^2} + 
{\cal O}\biggl(\frac{m_f^4}{T^4}\biggr)\biggr), 
\end{equation} 
here $m_f$ is the fermion mass and by $\mu$ we denoted the Debye mass.
All the other components of the polarization tensor
$\Pi_{\mu\nu}$ vanish in the zero-momentum limit
$k_0=0$, $\vec{k}\to 0$. 

Calculations of the CS term at finite temperature has drawn much
attention recently. 
In a number of papers (see, for instance~\cite{NieS:83,Popp:90,AiFZ:93}) 
there has been obtained by applying various perturbative
methods, that the one-loop expression for the CS mass is given at
finite temperature by
\begin{equation}\label{mT}
m(T)\sim m(T=0)\tanh\sfrac{m_f}{T}.
\end{equation}
Such  a smooth dependence on the temperature is, of course, in conflict 
with the gauge invariance. As has been argued in 
Refs.~\cite{Pis:87,DesGS:97}, since the CS mass has to take discrete 
values, it should not be modified by finite temperature corrections. 
Recently, there has been significant progress in resolving this 
puzzle. It has been found~\cite{DLL:97,FRS:97,AitF:98} that, 
when large gauge transformations are taken into account, the full 
effective action is gauge invariant, while the perturbative expansion 
is not. Thus, the apparent gauge non-invariance of the finite 
temperature Chern-Simons coefficient is only a result of considering the 
first term in the expansion of the effective action.\footnote{We will 
see below, that an effective mass appears in
our model, $\sim m^2+gT$, and the influence of CS mass temperature 
dependence of the form~(\ref{mT}) would be inessential at high $T$.}

Therefore, we can write the expression for the high-temperature
one-loop effective Lagrangian
\begin{equation}\label{LeffT}
{\cal L}_{eff} = -\frac{1}{4}G^{a}_{\mu\nu}G^{a}_{\mu\nu}
+ \frac{m}{4}\varepsilon^{\alpha\beta\gamma}\left(G^{a}_{\alpha\beta}
A^{a}_{\gamma}-\frac{g}{3}\varepsilon_{abc}A^{a}_{\alpha}
A^{b}_{\beta}A^{c}_{\gamma}\right)
+ \frac{1}{2}A_0^a\Pi_{00}(T)\delta^{ab}A_0^b,
\end{equation}
where $\Pi_{00}$ is given by the leading term in~(\ref{pi00}),
$\Pi_{00}=g^2T\frac{\ln 2}{2\pi}$.
The non-linear equations of motion derived from~(\ref{LeffT}) are
\begin{equation}\label{eqmot}
\partial_\nu G_{a}^{\nu\mu} + g\varepsilon^{abc}G_{b}^{\mu\nu}A_\nu^c
+ \frac{m}{2}\varepsilon^{\mu\alpha\beta}G^{a}_{\alpha\beta} 
+ g^{\mu 0}\Pi_{00}A_{0}^{a}= 0.
\end{equation}

Our aim is to find an exact solution to these equations. To this end
let us introduce the following ansatz for the gauge 
fields\footnote{This ansatz has been introduced in Ref.~\cite{Teh:90} in 
context of the search for an exact solution to Yang-Mills equations at 
zero temperature. The solutions found in that paper have infinite energy, 
and are time-dependent, $A_{a}^{\mu}\sim\exp(-t)$.}   
\begin{equation}\label{ansatz}
A_{i}^{a}=\hat\phi^a\hat\phi_i\Psi_1(\vec r\,)  
 + \delta_3^a \phi_i A(\vec r\,); \quad 
A_{0}^{a}=\hat\phi^a\Psi_2(\vec r\,).   
\end{equation}
Here and below we use the unit vectors of the cylindrical coordinate
system
$\hat\phi_i = \frac{\varepsilon_{ij}x^j}{r}$,
$\hat \varrho_i = \frac{x_i}{r}$.
Upon substituting the potentials~(\ref{ansatz}) into the 
system~(\ref{eqmot}), 
after rather lengthy but straightforward transformations the 
equations of motion take the form
\begin{eqnarray}\label{linsys}
- \partial^2_{r}\Psi_2 
- \frac{1}{r}\partial_{r}\Psi_2
- \frac{1}{r^2}\partial^2_{\varphi}\Psi_2 
- \frac{1}{r}\partial_{0}\partial_{\varphi}\Psi_1 
+ m\left(\partial_{r}\Psi_1 +\frac{1}{r}\Psi_1\right) 
+ \mu^2\Psi_2 = 0,                              
\nonumber\\
  \partial^2_{0}\Psi_1 
- \partial^2_{r}\Psi_1
- \frac{1}{r}\partial_{r}\Psi_1 
- \frac{2}{r^2}\partial^2_{\varphi}\Psi_1 
+ \frac{1}{r^2}\Psi_1 
- \frac{1}{r}\partial_{0}\partial_{\varphi}\Psi_2
+ m\partial_{r}\Psi_2 = 0,              \\
- \partial_{0}\partial_{r}\Psi_2
- \frac{1}{r}\partial_{r}\partial_{\varphi}\Psi_1
- \frac{1}{r^2}\partial_{\varphi}\Psi_1
+ m\left(\partial_0\Psi_1 
- \frac{1}{r}\partial_{\varphi}\Psi_2\right) = 0.
\nonumber
\end{eqnarray}
Here we have also set $A(r)=\frac{1}{gr}$. These equations are now 
linear, but still rather complicated for exhaustive analysis.
Further simplification can be achieved if we set the ansatz functions
$\Psi_1$, $\Psi_2$ to be time independent and axially symmetric. 
Under these assumptions our ansatz describes a static solution to
non-linear equations of motion that has the form of a vortex in both 
coordinate and internal spaces.
Then the third equation in~(\ref{linsys})
is satisfied identically, while the other two reduce to  
\begin{eqnarray}\label{linsys1}
- \partial^2_{r}\Psi_2 
- \frac{1}{r}\partial_{r}\Psi_2  
+ m\left(\partial_{r}\Psi_1 +\frac{1}{r}\Psi_1\right) 
+ \mu^2\Psi_2 = 0,                              
\nonumber \\
- \partial^2_{r}\Psi_1
- \frac{1}{r}\partial_{r}\Psi_1  
+ \frac{1}{r^2}\Psi_1 
+ m\partial_{r}\Psi_2 = 0.
\end{eqnarray}
Now one is able to solve the system exactly.
Multiplying the second equation by $m$ and taking the derivative
of the first one, we add the resultant expressions together to 
obtain
\begin{equation}\label{urBes}
   \Psi_2''' + \frac{1}{r}\Psi_2'' - 
      \left(\frac{1}{r^2} + M^2\right)\Psi_2' = 0,
\end{equation} 
here we denoted differentiation with respect to $r$ by prime, and
introduced the effective mass squared
\begin{equation}\label{Msq}
M^2 \equiv m^2 + \mu^2 = m^2 + g^2T\frac{\ln 2}{2\pi}. 
\end{equation} 
We see that both squared masses, Chern-Simons one and temperature 
generated, enter the equations with the same sign. Note that $M$ 
increases linearly as the temperature rises.
We recognize~(\ref{urBes}) as the modified 
Bessel equation for $\Psi_2'$ with the general solution:
\begin{equation}\label{dpsi1}
  \Psi_2'(r) = a K_1(Mr) + b I_1(Mr).
\end{equation}
Integrating~(\ref{dpsi1}) and leaving out the part of the
solution that diverges exponentially at spatial infinity, we obtain
\begin{equation}\label{psi2}
   \Psi_2(r) = - \frac{a}{M}K_0(Mr)
\end{equation}
and, solving equation for $\Psi_1(r)$,
\begin{equation}\label{psi1}
   \Psi_1(r) = \frac{am}{M^2} K_1(Mr) + \frac{c}{r}, 
\end{equation}
where $a$, $c$ are integration constants.

Finally, the expressions for the gauge potentials that solve the
non-linear equation of motion~(\ref{eqmot}) read  
\begin{eqnarray}\label{solut}
A_{i}^{a}&=&\hat\phi^a\hat\phi_i\left(\frac{am}{M^2} K_1(Mr) 
+ \frac{c}{r}\right)  
 + \delta_3^a \phi_i \frac{1}{gr}; 
\nonumber \\ 
A_{0}^{a} &=& - \hat\phi^a \frac{a}{M}K_0(Mr).
\end{eqnarray}
The properties of this vortex configuration will be discussed in the next 
section.

\section{The properties of the vortex solution}

Now let us investigate the physical characteristics of the
obtained field configuration. 
Electric field is along the radius in the coordinate space, while
in the  internal space it has only the angular component
\begin{equation}\label{elf}
  E_i^a = G^a_{i0} = 
  \hat\phi^a \hat\varrho_i \frac{\partial\Psi_2}{\partial r} =
  \hat\phi^a \hat\varrho_i a K_1(Mr). 
\end{equation}
Magnetic field is
\begin{equation}\label{maf}
  H^a = G^a_{12} = \frac{1}{2}\varepsilon^{ij}G^a_{ij} =  
  - \hat\phi^a (\frac{\partial\Psi_1}{\partial r} + \frac{1}{r}\Psi_1) 
  = \hat\phi^a \frac{a m}{M} K_0(Mr). 
\end{equation}
One can see that both electric and magnetic fields are screened
by the mass $M$, and vanish exponentially at spatial infinity.
This is what should be expected, for the Yang-Mills 
potentials~(\ref{solut}) asymptotically behave as
\begin{equation}\label{asym}
A_{i}^{a}\arrowcond{r\to\infty}{}\hat\phi^a\hat\phi_i\frac{c}{r}  
 + \delta_3^a\phi_i\frac{1}{gr}, 
\qquad 
A_{0}^{a}\arrowcond{r\to\infty}{}0, 
\end{equation}
which is a pure gauge.
Magnetic flux through the two-dimensional space is finite,
\begin{equation}\label{flux}
\Phi^a = \int d^2x H^a = \phi^a\frac{2\pi am}{M^3}.
\end{equation}

Now we turn to the total energy of the vortex determined as the
integral of the Hamiltonian density
\begin{eqnarray}\label{ham}
  {\cal E} &=& \int d^2x\,{\cal H} 
  = \int d^2x \biggl( A^{a}_{\sigma,0} 
     \frac{\partial{\cal L}}{\partial A^{a}_{\sigma,0}} 
  - {\cal L}\biggr) \\
  &=& 2\pi\int\! dr\,r 
  \Biggl[\frac{1}{2}\Bigl(\partial_r\Psi_2\Bigr)^2 
  + \frac{1}{2}\Bigl(\partial_r\Psi_1 + \frac{1}{r}\Psi_1\Bigr)^2
  + \frac{m}{2}\Bigl(\Psi_2\partial_r\Psi_1 - \Psi_1\partial_r\Psi_2
     + \frac{1}{r}\Psi_1\Psi_2\Bigr)
  - \frac{\mu^2}{2}\Psi_2^2 \Biggr]. \nonumber  
\end{eqnarray}
The energy density ${\cal H}$ in the above expression is divergent at 
the origin $r\to 0$
\begin{equation}\label{ham0}
 {\cal H}|_{r\to 0} = 
  \frac{1}{r^2}\left(\frac{a^2\mu^2}{2M^2}-\frac{a\,c\,m}{2M}\right) 
 + O(\ln r). 
\end{equation}
However, with appropriately chosen constants $a$, $c$ we can cancel
the term $\sim r^{-2}$ and make this singularity integrable. Then
the integrand in~(\ref{ham}) will read  
\begin{equation}\label{hamfin}
 {\cal H} = \frac{a^2\mu^2}{2M^2} K_1^2(Mr)
 - \frac{a^2\mu^2}{2M^3r} K_1(Mr)
 + \frac{a^2(2m^2-\mu^2)}{2M^2} K_1^2(Mr),
\end{equation}
and performing integration we get the following finite result for
the full energy of the vortex field configuration
\begin{equation}\label{ener}
  {\cal E} = \frac{a^2\pi(m^2 - \mu^2)}{(m^2 + \mu^2)^2}. 
\end{equation}
Thus, we have found the static solution to the Yang-Mills equations that 
possesses finite energy, i.e. a two-dimensional soliton.
Since asymptotic values of the fields correspond to the perturbative
vacuum solution, this solitonic configuration does not have a 
topological nature.
We note that the finiteness of energy became possible due
to the Debye mass being taken into account. If we set $\mu=0$, the only
choice of the constants leading to cancellation of $1/r^2$ divergence in
${\cal H}$ would be
$a=c=0$. This would leave us with a pure gauge solution equivalent to 
$A_{\mu}^{a}=0$. One can see as well, that the full energy changes 
sign: it is positive if $m>\mu$, and negative otherwise. 
Implications of this fact we will discuss below.

Now we consider the charge density of the system. The time component
of the relevant Noether current 
\begin{eqnarray}\label{char}
  gI_0^a &=& \partial^iE_i^a + mH^a - \mu^2A_0^a \\ 
  &=& \hat\phi^a(\partial^2_r\Psi_2 + \frac{1}{r}\partial_r\Psi_2
  - \mu^2\Psi_2 - m\partial_r\Psi_1 - \frac{m}{r}\Psi_1) = 0 
  \nonumber
\end{eqnarray}
vanishes on the vortex solution~(\ref{solut}).
We see that for all the calculated quantities their non-Abelian
structure factorizes as $\hat\phi^a$. This is true for the gauge
potentials as well, up to the pure gauge contribution. So, in a sense, 
the introduced ansatz can be thought of as an Abelian one.
Further, we calculate the electric charge density as $\mbox{div}\vec E$,
and integrate it over the whole space to obtain
\begin{equation}\label{qdist}
Q^a = \int d^2x\,\partial^i E_i^a = 
- 2\pi \phi^a\int dr\,r\,aMK_0(Mr) = -\phi^a\frac{2\pi a}{M}.
\end{equation}
So, the charge of the vortex appears to have the finite value.
On the other hand the electric field flux through the spatial
boundary vanishes due to the mass screening. These facts can 
be reconciled if we recall that the electric field behaves 
as $1/r$ at $r\to 0$
\begin{equation}\label{elf0}
  E_i^a|_{r\to 0} = 
  \hat\phi^a \hat\varrho_i \left(\frac{a}{Mr} + O(r\ln r)\right), 
\end{equation}
which is the 2-dimensional Coulomb law. Thus a point charge is 
present at the centre of the vortex, of the value 
\begin{equation}\label{point}
  Q_{point} = \frac{2\pi a}{M},
\end{equation}
and the total electric charge of the vortex $Q+Q_{point}$ is zero. 
It well corresponds to the picture described in~\cite{SVZ:94pr},
where it has been found that each lattice cell has a positive
core surrounded by negative charge distribution, in such a way
that the total charge vanishes.

\section{Discussion of the results}

In the present paper, the exact solution to the non-linear equations of 
motion for the topologically massive gauge field is derived at high 
temperature. To treat the problem consistently, the additional term 
$\sim A_0^2 \Pi_{00}(T)$ has been introduced to the effective Lagrangian. 
This is the key point of our analysis that has the general character at 
finite temperature in theories with the Chern-Simons mass. In this respect 
the situation is quite different as compared to the superconductivity and 
electroweak theory. In both those models there is no mixing between the 
electric and the magnetic sectors. The temperature influences the 
vortices through the mass of the condensed scalar field. In the considered
model the vortex is formed from the short-range colour magnetic and 
electric fields. The full energy of the configuration is finite. As the 
fields are screened by the effective mass $M$~(\ref{Msq}), at spatial 
infinity the perturbative vacuum solution is realized. The total charge of
the vortex is zero; the magnetic flux through the plane is finite.

Let us discuss the most important features of the discovered field
configuration. If we set $T=0$, our solution transforms into the 
divergent-energy solution of Ref.~\cite{Teh:90}. We stress once again that 
it is the addition of the Debye mass into the effective Lagrangian that 
makes the full energy finite. In fact, at high temperature it alone
determines the effective mass parameter $M\sim\mu$, and the role of
the CS mass is restricted to introducing the ansatz solutions to the 
non-linear field equations.    
It is worth noting that obtained configuration is composed from the 
gauge fields only. The vortex solutions in Chern-Simons theories 
obtained earlier~\cite{Kh:90} have been derived from the 
Lagrangians including Higgs fields. 

The physical contents of the derived field configuration varies.
As we see, the energy of the vortex~(\ref{ener}) is positive when 
$m^2>\mu^2\sim g^2T$. Thus, if this condition holds the derived solution 
describes the two-dimensional non-topological temperature soliton. 
Since in this case the perturbative approach used in Ref.~\cite{SZ:98e} 
is consistent, we have to conclude that when $T\leq m$ the ground state of 
the theory is the gluon condensate lattice with the period $\sim 1/m$ 
determined by the CS mass. The vortex solution, then, corresponds 
to the particle-like field excitation over the lattice vacuum structure.

At high temperature $\mu>m$ the situation is quite different. 
The vortex field configuration possesses negative energy and, therefore, 
one has to treat it as the ground state of the model. It is natural to 
consider this vacuum as the result of the evolution of the lattice 
vacuum structure of gluon condensate created at zero temperature, each
cell developing into a vortex.
Indeed, comparing the single lattice cell with the vortex puts forward
a number of similarities. In both cases we have gluon condensate formed
from the magnetic, as well as radial electric field. 
The magnetic field is maximal at the centre of the configuration, 
falling off towards the boundary; the magnetic flux is finite. 
The charge density distribution is similar as well: a core in the centre 
encircled by the charge of the opposite sign.
Thus, one can conjure up the following qualitative picture of the vacuum 
affected by the temperature. 
As the vacuum structure is heated, the period of the lattice, which is 
determined by the CS mass value, grows $d \sim 1/m \sim T$, as it follows 
from equation~(\ref{mT}). 
Eventually, the distance between flux `tubes' exceeds the characteristic 
scale~$\sim 1/M$, the lattice cells decouple and at high temperature 
each of them develops into the vortex field configuration. The ground
state thus becomes a plane covered with scattered non-interacting gauge 
field vortices.
The stability of the derived field configuration is ensured by the 
magnetic flux conservation.
Therefore, at high temperature the vacuum of the considered model is 
represented as the system of isolated vortices formed from colour electric 
and magnetic fields.   
 
%\bibliography{articles,translat}

\end{document}